\documentclass[draftcls,12pt,draftclsnofoot,onecolumn] {IEEEtran}
\usepackage{pifont}
\usepackage{stmaryrd}
\usepackage{mathrsfs}

\usepackage[cmex10]{amsmath}
\usepackage{cite}
\usepackage{amssymb}
\usepackage{amsmath}
\usepackage{amsthm}
\usepackage{mdwmath}
\usepackage{mdwtab}
\usepackage{array}
\usepackage{amssymb}
\usepackage{algorithmic}
\usepackage{graphicx}
\usepackage{fancyhdr}
\usepackage{algorithm} 
\usepackage{algorithmic} 

\makeatletter

\newcommand{\Rmnum}[1]{\expandafter\@slowromancap\romannumeral #1@}

\makeatother

\ifCLASSINFOpdf
\else
\fi
\begin{document}
%
\title{Cyclic Delay Transmission for Vector OFDM Systems}
%
%
%

\author{Weiliang Han,
        Xiang-Gen Xia,~\IEEEmembership{Fellow,~IEEE,}
        and~Jian-hua Ge
\thanks{Weiliang Han and Jian-hua Ge are with the State Key Laboratory of Integrated Service Networks, Xidian University, Xi'an, Shaanxi 710071, P.R. China, Han's work was done when he was visiting the University of Delaware (e-mail: wlhan@mail.xidian.edu.cn, jhge@xidian.edu.cn).}
\thanks{Xiang-Gen Xia is with Xidian University, Xian, China, and also the Department of Electrical and Computer Engineering, University of Delaware, Newark, DE 19716 USA (e-mail: xxia@ee.udel.edu).}
\thanks{Han and Ge's work were supported by the Program for Changjiang Scholars and Innovative Research Team in University (Grant No. IRT0852), the National Basic Research Program of China under Grant 2012CB316100, the “111” project (Grant No. B08038), the National Natural Science Foundation of China (Grant Nos. 61001207, 61101144 and 61101145). Xia's work was supported in part by the National Science Foundation (NSF) under Grant CCF-0964500.}
}

\maketitle
\vspace{-0.6in}
\enlargethispage{0.2in}
\begin{abstract}
Single antenna vector OFDM (V-OFDM) system has been proposed and investigated in the past. It contains the conventional OFDM and the single carrier frequency domain equalizer (SC-FDE) as two special cases and is flexible to choose any number of symbols in intersymbol interference (ISI) by choosing a proper vector size. In this paper, we develop cyclic delay diversity (CDD) transmission for V-OFDM when there are multiple transmit antennas (CDD-V-OFDM). Similar to CDD-OFDM systems, CDD-V-OFDM can also collect both spatial and multipath diversities. Since V-OFDM first converts a single input single output (SISO) ISI channel to a multi-input and multi-output (MIMO) ISI channel of order/length $K$ times less, where $K$ is the vector size, for a given bandwidth, the CDD-V-OFDM can accommodate $K$ times more transmit antennas than the CDD-OFDM does to collect all the spatial and multipath diversities. This property will specially benefit a massive MIMO system. We show that with the linear MMSE equalizer at each subcarrier, the CDD-V-OFDM achieves diversity order $d_{\text{CDD-V-OFDM}}^{\text{MMSE}} = \min \{ \lfloor 2^{-R}K \rfloor, N_t L \} +1$, where $R$ is the transmission rate, $N_t$ is the number of transmit antennas, and $L$ is the ISI channel length between each transmit and receive antenna pair. Simulations are presented to illustrate our theory.
\end{abstract}


%
\IEEEpeerreviewmaketitle

\section{Introduction}
\IEEEPARstart{O}{rthogonal} frequency division multiplexing (OFDM) systems have been well accepted  in broadband wireless communications, such as LTE \cite{3GPP}, WiMAX \cite{ieee2010ieee} and WiFi \cite{ieee2010ieee} systems. It is mainly because OFDM converts an ISI channel to multiple ISI free subchannels by using IDFT/DFT operations. When transmission data rates get higher and higher, signal bandwidths need to get wider and wider, which causes communication/ISI channel lengths longer and longer. A longer ISI channel forces the number $N$ of the subcarriers in an OFDM system larger in order to provide ISI free subchannels. A larger $N$ may cause more difficulties in OFDM implementations, such as high peak-to-average power ratio (PAPR), etc.

Single antenna V-OFDM is proposed in \cite{xia2000precoded, xia2001precoded}, which first converts a SISO ISI channel into an equivalent MIMO channel with memory, but the MIMO channel memory length (or order) is reduced by $K$ times compared to that of the SISO ISI channel, and then converts the equivalent MIMO channel with memory to multiple MIMO subchannels without memory. Since the MIMO channel length is $K$ times less, the CP length added to the V-OFDM can be also reduced by $K$ times compared to that of the OFDM for the SISO ISI channel. The cost for V-OFDM is that, although the multiple MIMO subchannels are constant vector/matrix channels and do not have memory, in each MIMO subchannel,  the $K$ information symbols  in each signal vector are still ISI channel. In other words, V-OFDM does not convert an ISI channel completely to multiple ISI free subchannels as OFDM does, but converts an ISI channel to multiple sub-channels with a fixed and flexible number, $K$,  of ISI symbols in each subchannel and $K$ can be arbitrarily chosen. As the two extreme cases of $K$, the conventional OFDM and single-carrier frequency domain equalizer (SC-FDE) correspond to the cases when $K=1$ and $N$, respectively. With V-OFDM, when an ISI channel length increases as its  bandwidth increases, the number $N$ of subcarriers may be fixed, while the vector size $K$ can be increased, which may be an alternative choice for a wide band system.

In wireless communications, an important technique to combat fading is to use multiple transmit antennas to collect spatial diversity \cite{grokop2009diversity, goldsmith2003capacity, kumar2007diversity}. For a broadband MIMO system,  there are two kinds of diversities, namely spatial diversity and multipath diversity. To collect both spatial and multipath diversities, space-frequency/time coding has been proposed in the literature to code information symbols across not only antennas but also subcarriers, see for example \cite{bolcskei2000space, su2003obtaining, zhang2007space, zhang2007full}, which, however, has a high decoding complexity at the receiver. A much simpler technique to collect both spatial and multipath diversities is the cyclic delay diversity (CDD) technique \cite{dammann2001standard, gore2002delay, dammann2002beamforming, dammann2002low, dammann2002performance, bossert2002cyclic, tan2004multicarrier, lodhi2005performance, bauch2006cyclic, mehana2012cyclic}, where the other transmit antennas transmit cyclically delayed versions of the signal transmitted at the first transmit antenna in every OFDM block. With CDD-OFDM, to collect the full spatial and multipath diversities, the product of the number of transmit antennas with the ISI channel length cannot be more than the number of subcarriers. When transmit antenna number is large, the full spatial and multipath diversities may not be collected by using the CDD technique when the number of subcarriers and the channel bandwidth are fixed.

In this paper, we propose a CDD transmission for V-OFDM (CDD-V-OFDM) for multiple transmit antenna systems. Similar to the CDD-OFDM, the other transmit antennas transmit cyclically delayed versions of the V-OFDM signals the first transmit antenna transmits in every V-OFDM block. Since an ISI channel length can be equivalently reduced by $K$ times in V-OFDM, with our proposed CDD-V-OFDM, for a fixed channel bandwidth (or channel length) and a fixed IDFT/DFT size (or number of subcarriers), CDD-V-OFDM can accommodate $K$ times more transmit antennas than CDD-OFDM does, where full spatial and multipath diversities can be collected. This property may benefit a massive MIMO system, where a massive number of transmit antennas are used. Or, if the number of transmit antennas and channel bandwidth are fixed, the IDFT/DFT size can be reduced by $K$ times with the CDD-V-OFDM that still collects the full spatial and multipath diversities, compared to that of the CDD-OFDM, which will consequently reduce the PAPR by $K$ times.

As mentioned earlier, a V-OFDM \cite{xia2000precoded, xia2001precoded, zhang2005synchronization, zhang2006iterative, han2010constellation, cheng2011v, li2012performance} converts an ISI channel to multiple constant matrix/vector channels, each of which has $K$ information symbols together. When $K$ is not small, the maximum-likelihood (ML) decoding of such a constant matrix/vector subchannel may be complex. In this paper, we investigate the minimum mean squared error (MMSE) equalizer for these subchannels. Following the results obtained in \cite{li2012performance}, we show that with the linear MMSE equalizer at each subcarrier, the CDD-V-OFDM achieves diversity order $d_{\text{CDD-V-OFDM}}^{\text{MMSE}} = \min \{ \lfloor 2^{-R}K \rfloor, N_t L \} +1$, where $R$ is the transmission rate, $N_t$ is the number of transmit antennas, and $L$ is the ISI channel length between each transmit and receive antenna pair.

This paper is organized as follows. In Section \Rmnum{2}, we first review OFDM and V-OFDM systems. In Section \Rmnum{3}, we derive a CDD-V-OFDM system, obtain an equivalent SISO ISI channel from the CDD-V-OFDM system for multiple transmit antennas, which also has an equivalent V-OFDM system. Then, we show the diversity order when linear MMSE equalizer is applied at each subchannel in the CDD-V-OFDM system. In Section IV, we present some simulation results to illustrate the analysis. In Section V, we conclude this paper.

\emph{Notations}: Throughout the paper. $ ()^H $ denotes the Hermitian conjugate transpose of a matrix or a vector. $ ()^T $ denotes the transpose. $ (())_N $ and $(())_{K}$ denote the modulo $N$ and the modulo $K$ operations. The bold face letters denote a vector or matrix. $\lceil \rceil$ and $\lfloor \rfloor$ denote the ceiling and the flooring operations, respectively.

\section{Brief Review of OFDM, and V-OFDM}
In this section, the conventional OFDM, and the V-OFDM system for single transmit antenna \cite{xia2000precoded, xia2001precoded} are briefly reviewed. The conventional OFDM system is shown in Fig. 1. $\textbf{x}=[x(0), x(1), ..., x(N-1)]^T$ is the frequency domain signal, $N$ is the number of the subcarriers in the OFDM system, and $\Gamma$ is the cyclic prefix length. After IDFT, the transmitted signal in time domain is
\begin{equation}\label{equ1_1_1}
  \overline{\textbf{x}} = [\overline{x}(0), ..., \overline{x}(N-1)]^T,
\end{equation}
and then the transmitted signal with CP is
\begin{equation}\label{equ1_1_2}
  \overline{\textbf{x}}_{CP} = [\overline{x}(N-\Gamma), ..., \overline{x}(N-1), \overline{x}(0), ..., \overline{x}(N-1)]^T.
\end{equation}
\begin{figure}
  \centering
  \includegraphics[width=0.6\textwidth,angle=270]{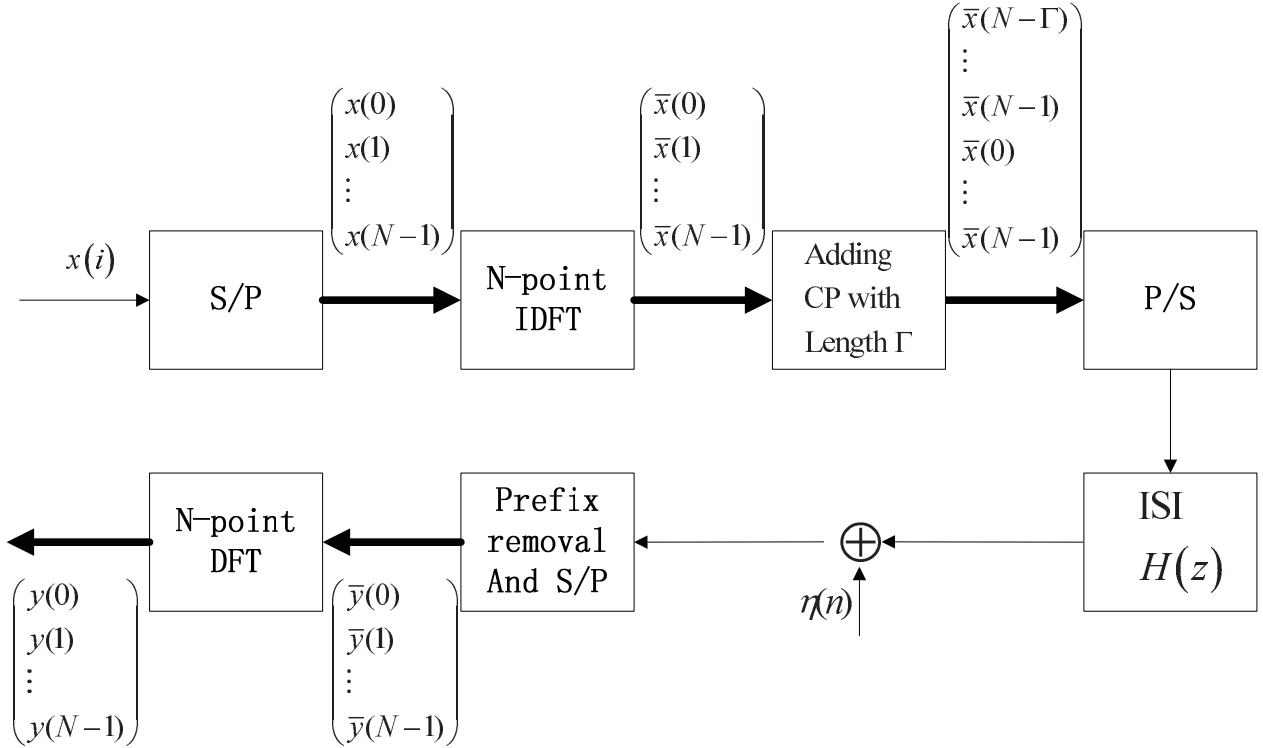}\\ %
  \caption{Conventional OFDM system}\label{Fig1}
\end{figure}
Assume that the ISI channel disperse has the following transfer function:
\begin{equation}\label{equ1_1_3}
  H(z) = \sum\limits_{l=0}^{N-1}z^{-l} h(l)
\end{equation}
and
\begin{align}\label{equ1_1_4}
 \textbf{h}
 &= [h(0), h(1), ..., h(N-1)] \nonumber \\
 &= [h(0), h(1), ..., h(L-1), 0, ..., 0]
\end{align}
is the corresponding channel impulse response (CIR) in time domain. At the receiver, CP is removed first, and we get
\begin{equation}\label{equ1_1_5}
  \overline{\textbf{y}} = [\overline{y}(0), \overline{y}(1), ..., \overline{y}(N-1)]^T.
\end{equation}
Then, transfer the received signal to frequency domain by DFT with size $N$, we obtain
\begin{equation}\label{equ1_1_6}
  \textbf{y} = [y(0), y(1), ..., y(N-1)]^T.
\end{equation}

\begin{figure}
  \centering
  \includegraphics[width=0.6\textwidth,angle=270]{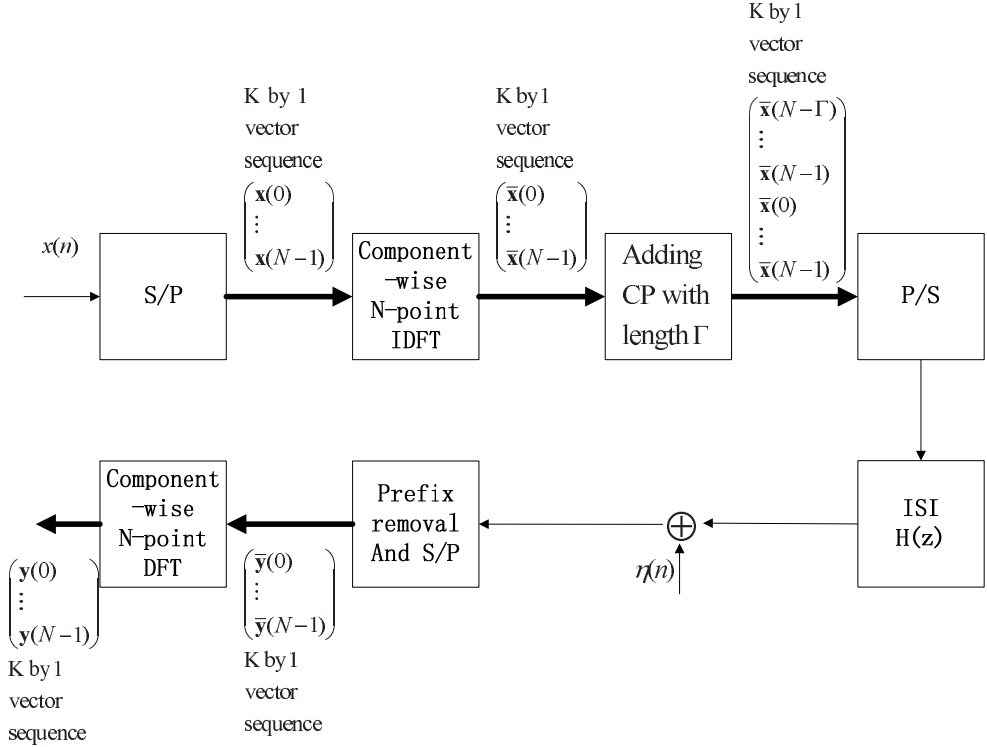}\\
  \caption{Vector OFDM system}\label{Fig2}
\end{figure}

Fig. \ref{Fig2} illustrates the block diagram of the V-OFDM system. Data block $\textbf{x} = [x(0), x(1), ..., x(KN-1)]$ is transmitted for each V-OFDM block, which contains $N$ vectors with $K$ symbols in each vector. (If we keep the block length $N$ unchanged, the block can be divided into small vectors with low PAPR and save CP, here we use $KN$ as the block length). Let
\begin{equation}\label{equ1_1_7}
  \textbf{x} = [\textbf{x}(0)^T, \textbf{x}(1)^T, ..., \textbf{x}(N-1)^T]^T,
\end{equation}
where $\textbf{x}(n) = [x_0(n), x_1(n), ..., x_{K-1}(n)]^T , n=0,1,...N-1$, and $x_k(n) = x(k+nK)$. V-OFDM does component-wise  IDFT of size $N$ over the vectors, i.e.,
\begin{align}\label{equ1_1_8}
  \overline{x}_k(q) = \frac{1}{\sqrt{N}} \sum\limits_{n=0}^{N-1} x_k(n) exp(j \frac{2\pi nq}{N}), q=0,1,...N-1,
\end{align}
we have the signal in time domain as
\begin{align}\label{equ1_1_9}
  \overline{\textbf{x}} = [\overline{\textbf{x}}(0)^T, \overline{\textbf{x}}(1)^T, ..., \overline{\textbf{x}}(N-1)^T]^T,
\end{align}
Then, the transmitted symbol sequence after adding cyclic prefix (CP) in time domain can be written as
\begin{align}\label{equ1_1_10}
  \overline{\textbf{x}}_{CP} = [\overline{\textbf{x}}^T(N-\Gamma), ..., \overline{\textbf{x}}^T(N-1), \overline{\textbf{x}}^T(0), ..., \overline{\textbf{x}}^T(N-1)]^T,
\end{align}
where $\overline{\textbf{x}}(n) = [\overline{x}_0(n),\overline{x}_1(n),...,\overline{x}_{K-1}(n)]^T, n=0,1,...N-1$, and $\overline{x}_k(q) = \overline{x}(k+qK)$. After removing the CP at the receiver, we have the signal in time domain as
\begin{equation}\label{equ1_1_11}
  \overline{\textbf{y}} = [\overline{\textbf{y}}(0)^T, \overline{\textbf{y}}(1)^T, ..., \overline{\textbf{y}}(N-1)^T]^T,
\end{equation}
where $\overline{\textbf{y}}(n) = [\overline{y}_{0}(n), \overline{y}_{1}(n), ..., \overline{y}_{K-1}(n)]^T$. With component-wise DFT in the similar way as (\ref{equ1_1_8}), the signal will be transformed to frequency domain as
\begin{equation}\label{equ1_1_12}
  \textbf{y} = [\textbf{y}(0)^T, \textbf{y}(1)^T, ..., \textbf{y}(N-1)^T]^T,
\end{equation}
where $\textbf{y}(n) = [y_{0}(n), y_{1}(n), ..., y_{K-1}(n)]^T$. The relationship between the transmitted and received signals in z domain can be expressed as follows. From \cite{xia2001precoded}, assume that $\overline{x}_k(z)$ and $\overline{y}_k(z)$ are the $k$th polyphase components of the z-transforms of the transmitted and received signals, respectively. Then, the received signal can be written in the following MIMO form in $z$ domain
\begin{equation}\label{equ1_1_13}
  \overline{\textbf{y}}(z) = \mathcal{H}_{V-OFDM}(z) \overline{\textbf{x}}(z) + \xi(z),
\end{equation}
where $\overline{\textbf{y}}(z) = [\overline{y}_0(z), \overline{y}_1(z), ..., \overline{y}_{K-1}(z)]^T$, $\overline{\textbf{x}}(z) = [\overline{x}_0(z), \overline{x}_1(z), ..., \overline{x}_{K-1}(z)]^T$, $\xi(z)$ is the $z$ transform of the noise,
\begin{align}\label{equ1_1_14}
  \mathcal{H}_{V-OFDM}(z)
  =
  \begin{bmatrix}
  H_0(z)     & z^{-1}H_{K-1}(z) & \cdots & z^{-1}H_{1}(z) \\
  H_1(z)     & H_0(z)           & \cdots & z^{-1}H_{2}(z) \\
  \vdots     & \vdots           & \vdots & \vdots \\
  H_{K-2}(z) & H_{K-3}(z)       & \cdots & z^{-1}H_{K-1}(z) \\
  H_{K-1}(z) & H_{K-2}(z)       & \cdots & H_0(z)
  \end{bmatrix},
\end{align}
and
\begin{equation}\label{equ1_1_15}
  H_{k}(z) = \sum\limits_{q=0}^{N-1} h_k(q)z^{-q}, k=0,1,...,K-1,
\end{equation}
where
\begin{align}\label{equ1_1_16}
\textbf{h}_k &= [h_k(0), h_k(1),..., h_k(N-1)] \nonumber \\
&\triangleq [h_k(0), h_k(1),..., h_k(\lceil L/K \rceil-1), 0,..., 0] \nonumber \\
&= [h(k), h(k+K),..., h(k+(\lceil L/K \rceil-1)K), 0,..., 0]
\end{align}
is the $k$th polyphase component of (\ref{equ1_1_4}) and (\ref{equ1_1_16}) is the $k$th polyphase component of (\ref{equ1_1_3}). Note that, for the convenience later, we use length $N$ for the $k$th polyphase component $h_k(l)$ in time domain as well but its number of non-zero components is at most $\lceil L/K \rceil$. It is clear that (\ref{equ1_1_16}) has the same form as the conventional OFDM system (\ref{equ1_1_4}) for each $k$, while the channel length is at most $\lceil L/K \rceil$ as shown in (\ref{equ1_1_16}).

\section{Proposed CDD-V-OFDM system}
In this section, we first describe our proposed CDD-V-OFDM system and present some of its properties. We then present its performance analysis when the linear MMSE equalizer/receiver is used for every subchannel. Also, for convenience, in this section, we assume $N=KN_0$ for some positive integer $N_0$. A general case of $N$ can be similarly studied but the notations become more tedious.
\subsection{Proposed System Description}
\begin{figure}
  \centering
  \includegraphics[width=0.65\textwidth,angle=270]{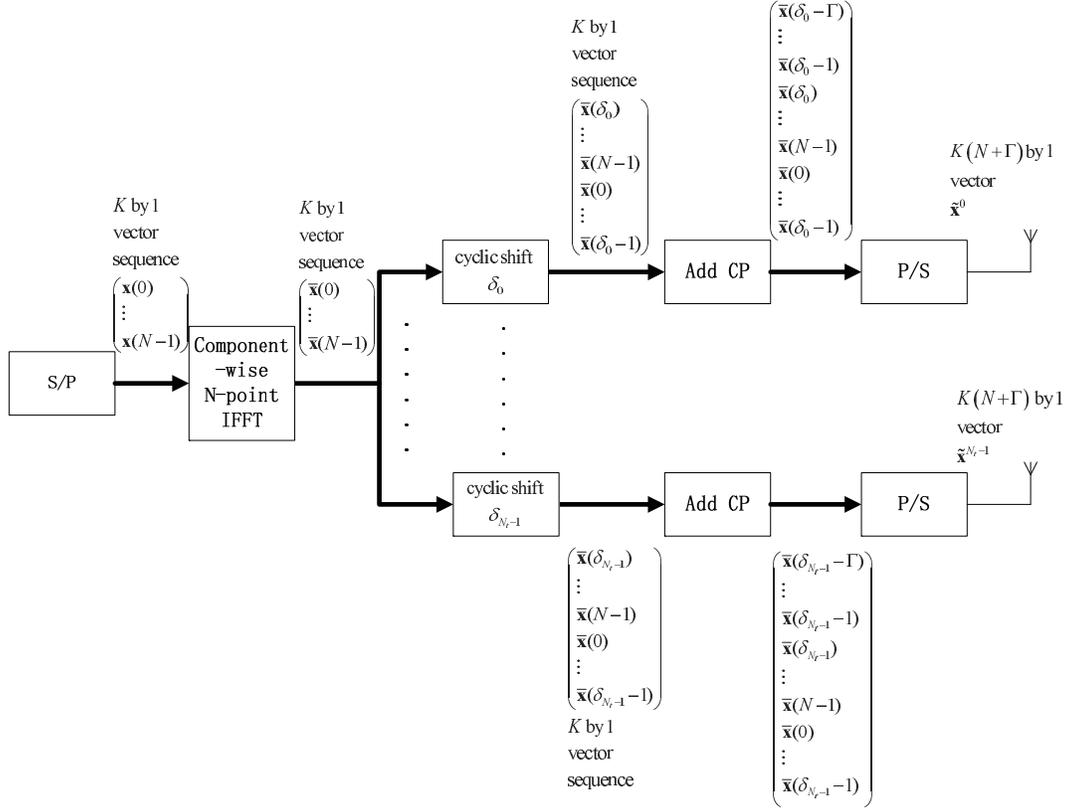}\\
  \caption{CDD-V-OFDM transmission}\label{Fig3}
\end{figure}

Combining V-OFDM with CDD transmission can enhance the diversity order and accommodate more antennas in the system, especially in massive MIMO system as we shall see in this section. Our proposed CDD-V-OFDM system is shown in Fig. \ref{Fig3}. In this section, the CDD-V-OFDM system and its equivalent vectorized MIMO ISI CIR are presented, where a single receive antenna is used for convenience.

Assume the original data sequence as $[x(0), x(1), ..., x(KN-1)]$, which is blocked into $N$ vectors with $K$ symbols in each vector:
\begin{equation}\label{equ1_2_1}
  \textbf{x}=[(\textbf{x}(0))^T, (\textbf{x}(1))^T, ..., (\textbf{x}(N-1))^T]^T,
\end{equation}
where $\textbf{x}(n) = [x_0(n), x_1(n), ..., x_{K-1}(n)]$ and $x_k(n) = x(k+nK)$, $k=0, 1, ..., K-1$. After the  component-wise IDFT shown in (\ref{equ1_1_8}), the signal in time domain is
\begin{equation}\label{equ1_2_2}
  \overline{\textbf{x}} = [(\overline{\textbf{x}}(0))^T, (\overline{\textbf{x}}(1))^T, ... (\overline{\textbf{x}}(N-1))^T]^T,
\end{equation}
where $\overline{\textbf{x}}(q) = [\overline{x}_0(q), ..., \overline{x}_{K-1}(q)]$, and $\overline{x}_k(q) = \overline{x}(k+qK)$, $k=0, 1, ..., K-1$. There are $N_t$ transmit antennas in this system and let us consider the $m$th transmit antenna. The cyclic shift with amount $\delta_m$ in the $m$th transmit antenna leads the signal in (\ref{equ1_2_2}) to
\begin{align}\label{equ1_2_3}
  [(\overline{\textbf{x}}^m(0))^T,...,(\overline{\textbf{x}}^m(N-1))^T] = [(\overline{\textbf{x}}(\delta_m))^T,...,(\overline{\textbf{x}}(N-1))^T, (\overline{\textbf{x}}(0))^T, ..., (\overline{\textbf{x}}(\delta_m-1))^T]^T
\end{align}
where
\begin{align}\label{equ1_2_4}
  \overline{\textbf{x}}^m(q)
  &= [\overline{x}_0^m(q), \overline{x}_1^m(q), ..., \overline{x}_{K-1}^m(q)]^T \nonumber \\
  &= [\overline{x}_0((q+\delta_m))_N, \overline{x}_1((q+\delta_m))_N, ..., \overline{x}_{K-1}((q+\delta_m))_N]^T,
\end{align}
where $q = 0,..., N-1$. Then, the transmit signal after appending CP for the $m$th transmit antenna is
\begin{align}\label{equ1_2_5}
  \widetilde{\textbf{x}}^m
  &= [\widetilde{\textbf{x}}^m(0))^T, ..., (\widetilde{\textbf{x}}^m(\Gamma-1))^T, (\widetilde{\textbf{x}}^m(\Gamma))^T,...,(\widetilde{\textbf{x}}^m(N+\Gamma-1))^T]^T \nonumber \\
  &\triangleq [\overbrace{(\overline{\textbf{x}}^m(N-\Gamma))^T, ..., (\overline{\textbf{x}}^m(N-1))^T, (\overline{\textbf{x}}^m(0))^T,...,(\overline{\textbf{x}}^m(N-1))^T}^{(N+\Gamma) \text{vectors}}]^T
\end{align}
where $\widetilde{\textbf{x}}^m(q) = [\widetilde{x}_0^m(q), \widetilde{x}_1^m(q), ..., \widetilde{x}_{K-1}^m(q)]^T$ and $\widetilde{x}_k^m(q) = \widetilde{x}^m(k+Kq), q = 0,1,...,N+\Gamma-1$.
Consider the original CIR from the $m$th antenna as
\begin{align}\label{equ1_2_6}
\textbf{h}^m &= [h^m(0), h^m(1), ..., h^m(N-1)] \nonumber \\
 &= [h^m(0), ..., h^m(L-1), 0, ..., 0].
\end{align}
Then, the $k$th polyphase of the received signal from the $m$th antenna can be expressed as a linear convolution of $\widetilde{\textbf{x}}^m$ and the original CIR from the $m$th antenna as
\begin{align}\label{equ1_2_7}
  \overline{y}_k^m(q)
  &= \overline{y}^m(Kq+k) \nonumber \\
  &= \sum\limits_{l=0}^{N-1} h^m(l) \widetilde{x}^m(Kq+k-l).
\end{align}
Referring to the V-OFDM, we need to transform the linear convolution between the CIR and the transmit signals to the cyclic convolution between the vectorized ISI channel and the vectorized OFDM signals. Then, we do the same vectorization operation for the ISI channel as the V-OFDM signal. Assume that $\lfloor l/K \rfloor =t$, and $((l))_K = s$, for $0 \leq k \leq K-1$, we have $((k-s))_K = 0, 1, ..., K-1$ and $\lfloor (k-s)/K \rfloor$ is
\begin{equation}\label{equ1_2_8}
  \lfloor (k-s)/K \rfloor =
  \begin{cases}
  0 & \text{if } k \geq s \\
  -1& \text{if } k < s.
  \end{cases}
\end{equation}
Then, the index of the signal $\widetilde{x}^m$ in (\ref{equ1_2_7}) is separated into two parts: one is the index of a symbol in a vector, i.e., $((k-s))_K$, and the other is the index of the vector, i.e., $(q-t+\lfloor (k-s)/K \rfloor)$. Substituting $l$ with $s+tK$, (\ref{equ1_2_7}) is changed to
\begin{align}\label{equ1_2_9}
  \overline{y}_k^m(q)
  = \sum\limits_{s=0}^{K-1} \sum\limits_{t=0}^{N_0-1} h^m(s+tK) \widetilde{x}^m(((k-s))_K +K(q-t+\lfloor (k-s)/K \rfloor)).
\end{align}
Consider the first equation in (\ref{equ1_2_5}), the transmit signal $\widetilde{\textbf{x}}$ consists of $N+\Gamma$ vectors with $K$ symbols in each vector. Thus, (\ref{equ1_2_9}) is changed to
\begin{align}\label{equ1_2_10}
  \overline{y}_k^m(q)
  = \sum\limits_{s=0}^{K-1} \sum\limits_{t=0}^{N_0-1} h^m(s+tK) \widetilde{x}_{((k-s))_K}^m(q-t+\lfloor (k-s)/K \rfloor),
\end{align}
where $k = 0, ..., K-1$ and $q = 0, ..., N+\Gamma-1$. Inside $\widetilde{x}_{k}^m(q)$, $q = 0, ..., N+\Gamma-1$, the region of CP is $q = 0, ...,\Gamma-1$, which means, $\widetilde{x}_{k}^m(q) = \widetilde{x}_{k}^m(q+N)$, for $q = 0, ...,\Gamma-1$. After removing the CP at the receiver, the region $q = [\Gamma, ..., N+\Gamma-1]$ is what we concern in the received signal. Here we divide the summation period of $t$ into two parts, $q-t+\lfloor (k-s)/K \rfloor \geq \Gamma$ and $q-t+\lfloor (k-s)/K \rfloor < \Gamma$. Then, we have
\begin{align}\label{equ1_2_11}
    \overline{y}_k^m(q)
    &= \sum\limits_{s=0}^{K-1} \sum\limits_{t=0}^{N_0-1} h^m(s+tK) \widetilde{x}_{((k-s))_K}^m(q-t+\lfloor (k-s)/K \rfloor) \nonumber \\
    &= \sum\limits_{s=0}^{K-1} \Big\{ \sum\limits_{t=0}^{q - \Gamma +\lfloor (k-s)/K \rfloor} h^m(s+tK) \widetilde{x}_{((k-s))_K}^m(q-t+\lfloor (k-s)/K \rfloor) \nonumber \\
    &+ \sum\limits_{t=q - \Gamma +\lfloor (k-s)/K \rfloor+1}^{N_0-1} h^m(s+tK) \widetilde{x}_{((k-s))_K}^m(q-t+\lfloor (k-s)/K \rfloor)\Big\}
\end{align}
Because $\widetilde{x}_{k}^m(q) = \widetilde{x}_{k}^m(q+N)$, for $q = 0, ...,\Gamma-1$, the second item of  (\ref{equ1_2_10}) is changed to
\begin{align}\label{equ1_2_12}
    \overline{y}_k^m(q)
    &= \sum\limits_{s=0}^{K-1} \Big\{ \sum\limits_{t=0}^{q-\Gamma+\lfloor (k-s)/K \rfloor} h^m(s+tK) \widetilde{x}_{((k-s))_K}^m(q-t+\lfloor (k-s)/K \rfloor) \nonumber \\
    &+ \sum\limits_{t=q-\Gamma+\lfloor (k-s)/K \rfloor+1}^{N_0-1} h^m(s+tK) \widetilde{x}_{((k-s))_K}^m(q-t+N+\lfloor (k-s)/K \rfloor)\Big\}
\end{align}
From (\ref{equ1_2_4}) we know $\overline{x}_{k}^m(q-\Gamma) = \widetilde{x}_{k}^m(q)$, for $q = \Gamma, ..., N+\Gamma-1$, then we have
\begin{align}\label{equ1_2_13}
    \overline{y}_k^m(q) &= \sum\limits_{s=0}^{K-1} \Big\{ \sum\limits_{t=0}^{q-\Gamma+\lfloor (k-s)/K \rfloor} h^m(s+tK) \overline{x}_{((k-s))_K}^m(q-t+\lfloor (k-s)/K \rfloor-\Gamma) \nonumber \\
    &+ \sum\limits_{t=q-\Gamma+\lfloor (k-s)/K \rfloor+1}^{N_0-1} h^m(s+tK) \overline{x}_{((k-s))_K}^m(q-t+N+\lfloor (k-s)/K \rfloor-\Gamma)\Big\}
\end{align}
Substituting $q-\Gamma$ with $q$, we have
\begin{align}\label{equ1_2_14}
    \overline{y}_k^m(q+\Gamma)
    &= \sum\limits_{s=0}^{K-1} \Big\{ \sum\limits_{t=0}^{q+\lfloor (k-s)/K \rfloor} h^m(s+tK) \overline{x}_{((k-s))_K}^m(q-t+\lfloor (k-s)/K \rfloor) \nonumber \\
    &+ \sum\limits_{t=q+\lfloor (k-s)/K \rfloor+1}^{N_0-1} h^m(s+tK) \overline{x}_{((k-s))_K}^m(q-t+N+\lfloor (k-s)/K \rfloor)\Big\} \nonumber \\
    &= \sum\limits_{s=0}^{K-1} \Big\{ \sum\limits_{t=0}^{q+\lfloor (k-s)/K \rfloor} h^m(s+tK) \overline{x}_{((k-s))_K}^m \left( ((q-t+\lfloor (k-s)/K \rfloor))_N \right) \nonumber \\
    &+ \sum\limits_{t=q+\lfloor (k-s)/K \rfloor+1}^{N_0-1} h^m(s+tK) \overline{x}_{((k-s))_K}^m \left( ((q-t+\lfloor (k-s)/K \rfloor))_N \right) \Big\} \nonumber \\
    &= \sum\limits_{s=0}^{K-1} \sum\limits_{t=0}^{N_0-1} h^m(s+tK) \overline{x}_{((k-s))_K}^m \left( ((q-t+\lfloor (k-s)/K \rfloor))_N  \right)
\end{align}
where $q = 0, ..., N-1$. Then, the $k$th polyphase component of the received signal can be expressed by the cyclic convolution of the vectorized OFDM signal and the vectorized ISI CIR as
\begin{align}\label{equ1_2_15}
  \overline{y}_k^m(q+\Gamma)
  =
  \sum\limits_{s=0}^{K-1} \sum\limits_{t=0}^{N_0-1} h^m(s+tK) \overline{x}_{((k-s))_K}^m \left( ((q-t
  +\lfloor (k-s)/K \rfloor))_N \right)
\end{align}
where $((n))_N$ denotes $m$ modulo $N$ operation and $q = 0, ..., N-1$. For the cyclic shift in (\ref{equ1_2_2}) and (\ref{equ1_2_3}), we have $\overline{x}_k^m(q) = \overline{x}_k((q+\delta_m))_N$. Then, (\ref{equ1_2_14}) is rewritten as
\begin{align}\label{equ1_2_16}
  \overline{y}_k^m(q+\Gamma)
  &= \sum\limits_{s=0}^{K-1} \sum\limits_{t=0}^{N_0-1} h^m(s+tK) \overline{x}_{((k-s))_K} \left( ((q-t+\lfloor (k-s)/K \rfloor+\delta_m))_N \right) \nonumber \\
  &= \sum\limits_{s=0}^{K-1} \sum\limits_{t=0}^{N-1} h_s^m(t) \overline{x}_{((k-s))_K} \left( ((q-t+\lfloor (k-s)/K \rfloor+\delta_m))_N \right)
\end{align}
where, similar to (\ref{equ1_1_16}), we have
\begin{align}\label{equ1_2_17}
    \textbf{h}_s^m = &[h_s^m(0), h_s^m(1), ..., h_s^m(N-1)] \nonumber \\
    \triangleq &[h_s^m(0), h_s^m(1),..., h_s^m(\lceil L/K \rceil-1), 0..., 0] \nonumber \\
    = &[h^m(s), h^m(s+K),..., h^m(s+(\lceil L/K \rceil-1)K), 0..., 0].
\end{align}
Substituting $t-\delta_m - \lfloor (k-s)/K \rfloor$ with $t'$, and following the similar process as (\ref{equ1_2_11})-(\ref{equ1_2_17}), for $q=0,1,...,N-1$, we obtain
\begin{align}\label{equ1_2_18}
    y_k^m(q+\Gamma)
    &= \sum\limits_{s=0}^{K-1} \sum\limits_{t'=-\delta_m-\lfloor (k-s)/K \rfloor}^{N-1-\delta_m-\lfloor (k-s)/K \rfloor} h_s^m \left(t'+\lfloor (k-s)/K \rfloor+\delta_m \right) \overline{x}_{((k-s))_K} \left( ((q-t'))_N \right) \nonumber \\
    &= \sum\limits_{s=0}^{K-1} \{ \sum\limits_{t'=-\delta_m-\lfloor (k-s)/K \rfloor}^{-1} h_s^m \left(t'+\lfloor (k-s)/K \rfloor+\delta_m \right) \overline{x}_{((k-s))_K} \left( ((q-t'))_N \right) \nonumber \\
    &+ \sum\limits_{t'=0}^{N-1-\delta_m-\lfloor (k-s)/K \rfloor} h_s^m \left(t'+\lfloor (k-s)/K \rfloor+\delta_m \right) \overline{x}_{((k-s))_K} \left( ((q-t'))_N \right) \}\nonumber \\
    &= \sum\limits_{s=0}^{K-1} \{ \sum\limits_{t'=N-\delta_m-\lfloor (k-s)/K \rfloor}^{N-1} h_s^m \left(t'+\lfloor (k-s)/K \rfloor+\delta_m-N \right) \overline{x}_{((k-s))_K} \left( ((q-t'))_N \right) \nonumber \\
    &+ \sum\limits_{t'=0}^{N-1-\delta_m-\lfloor (k-s)/K \rfloor} h_s^m \left(t'+\lfloor (k-s)/K \rfloor+\delta_m \right) \overline{x}_{((k-s))_K} \left( ((q-t'))_N \right) \}\nonumber \\
    &= \sum\limits_{s=0}^{K-1} \{ \sum\limits_{t'=N-\delta_m-\lfloor (k-s)/K \rfloor}^{N-1} h_s^m \left(((t'+\lfloor (k-s)/K \rfloor+\delta_m))_{N} \right) \overline{x}_{((k-s))_K} \left( ((q-t'))_N \right) \nonumber \\
    &+ \sum\limits_{t'=0}^{N-1-\delta_m-\lfloor (k-s)/K \rfloor} h_s^m \left(((t'+\lfloor (k-s)/K \rfloor+\delta_m))_{N} \right) \overline{x}_{((k-s))_K} \left( ((q-t'))_N \right) \} \nonumber \\
    &= \sum\limits_{s=0}^{K-1} \sum\limits_{t'=0}^{N-1} h_s^m(((t'+\lfloor (k-s)/K \rfloor + \delta_m))_N) \overline{x}_{((k-s))_K} \left( ((q-t'))_N \right)
\end{align}
Since (\ref{equ1_2_18}) is the signal from the $m$th antenna, the overall received signal is the sum of the signals from all the antennas, which is
\begin{align}\label{equ1_2_19}
\overline{y}_k(q+\Gamma)
&= \sum\limits_{m=0}^{N_t-1} \overline{y}_k^m(q+\Gamma) \nonumber \\
&= \sum\limits_{s=0}^{K-1} \sum\limits_{t'=0}^{N-1} \sum\limits_{m=0}^{N_t-1} h_s^m(((t'+\lfloor (k-s)/K \rfloor + \delta_m))_N) \overline{x}_{((k-s))_K} \left( ((q-t'))_N \right) \nonumber \\
&= \sum\limits_{s=0}^{K-1} \sum\limits_{t'=0}^{N-1} h_{eqv_s}(((t'+\lfloor (k-s)/K \rfloor))_N) \overline{x}_{((k-s))_K} \left( ((q-t'))_N \right)
\end{align}
where
\begin{align}\label{equ1_2_20}
  &h_{eqv_s}(((t'+\lfloor (k-s)/K \rfloor))_N) \nonumber \\
  &\triangleq \sum\limits_{m=0}^{N_t-1} h_s^m(((t'+\lfloor (k-s)/K \rfloor + \delta_m))_N).
\end{align}
Equation (\ref{equ1_2_19}) consists of $K$ polyphase components as $s=0,1,...,K-1$. The equivalent SISO channel over all the polyphase components and all the antennas is
\begin{align}\label{equ1_2_21}
    \textbf{h}_{eqv} \triangleq & [h_{eqv}(0), h_{eqv}(1), ..., h_{eqv}(NK-1)] \nonumber \\
    = & [h_{eqv_0}(0), h_{eqv_1}(0), ..., h_{eqv_{K-1}}(0), \nonumber \\
    & h_{eqv_0}(1), h_{eqv_1}(1), ..., h_{eqv_{K-1}}(1), \nonumber \\
    & ... \nonumber \\
    & h_{eqv_0}(N-1), h_{eqv_1}(N-1), ..., h_{eqv_{K-1}}(N-1)].
\end{align}
Then, the $k$th polyphase component of length $N$ in $\textbf{h}_{eqv}$ is
\begin{align}\label{equ1_2_22}
    \textbf{h}_{eqv_k} &\triangleq [h_{eqv_k}(0), h_{eqv_k}(1), ..., h_{eqv_k}(N-1)] \nonumber \\
                       &= [h_{eqv}(k), h_{eqv}(k+K), ..., h_{eqv}(k+(N-1)K)].
\end{align}
Comparing (\ref{equ1_1_16}) and (\ref{equ1_2_22}), we can see that the equivalent ISI channel of the CDD-V-OFDM has the same form as that of V-OFDM for a SISO ISI channel, while in the CDD-V-OFDM system, the equivalent ISI channel is the sum of the $N_t$ cyclically shifted channels  from different transmit antennas. Next, the vectorized ISI channel has the z-domain pseudocirculant matrix form as follows
\begin{align}\label{equ1_2_23}
  \mathcal{H}_{eqv}(z) =
  \begin{bmatrix}
  H_{eqv_0}(z)     & z^{-1}H_{eqv_{K-1}}(z) & \cdots & z^{-1}H_{eqv_1}(z) \\
  H_{eqv_1}(z)     & H_{eqv_0}(z)           & \cdots & z^{-1}H_{eqv_2}(z) \\
  \vdots           & \vdots                 & \vdots & \vdots \\
  H_{eqv_{K-2}}(z) & H_{eqv_{K-3}}(z)       & \cdots & z^{-1}H_{eqv_{K-1}}(z) \\
  H_{eqv_{K-1}}(z) & H_{eqv_{K-2}}(z)       & \cdots & H_{eqv_0}(z)
  \end{bmatrix},
\end{align}
where $H_{eqv_k}(z) = \sum\limits_{t'=0}^{N -1}h_{eqv_k}(t') z^{-t'}$, $k = 0, 1, ..., K-1$, is the z-transform of (\ref{equ1_2_22}), which is the $k$th polyphase component of (\ref{equ1_2_21}). From \cite{mehana2012cyclic} and \cite{li2012performance}, the diversity order is closely related to the length of a non-overlapping multipath channel. To achieve the maximal spatial and multipath diversities, there should be no channel overlap between the CIRs from different antennas. Thus, the cyclic shift amount should meet the following condition, if $N \geq N_t L/K$,
\begin{equation}\label{equ1_2_24}
    \delta_m = (m-1)\delta, \delta \in [L/K, N/N_t].
\end{equation}
Then, the equivalent ISI SISO channel becomes
\begin{align}\label{equ1_2_25}
  \textbf{h}_{eqv} =
  \{ &[h^0(0),h^0(1)...,h^0(L-1), \underbrace{0,...,0}_{\delta-L}], [h^{1}(0),h^{1}(1)..., h^{1}(L-1), \underbrace{0, ..., 0}_{\delta-L}],\nonumber \\
   &..., [h^{N_t-1}(0),h^{N_t-1}(1)..., h^{N_t-1}(L-1), \underbrace{0, ...,0}_{\delta-L}] ,0, ..., 0 \}.
\end{align}

The equivalent ISI channel of the CDD-V-OFDM can also be regarded as an equivalent ISI channel in CDD-OFDM. The original CIR from the $m$th antenna in CDD-OFDM system is the same as (\ref{equ1_2_5}). Then, the overall equivalent ISI channel of the CDD-OFDM is
\begin{equation}\label{equ1_2_26}
  \textbf{h} = [h_{eqv}(0), h_{eqv}(1), ..., h_{eqv}(N-1)],
\end{equation}
where $h_{eqv}(t) = \sum\limits_{m=0}^{N_t-1} h^m \left( ((t+\delta_m))_{N} \right)$. When the condition $\delta_m \in [L,N/N_t]$ with $N \geq N_t L$ holds, there are no overlaps between the CIRs from different antennas, and the equivalent ISI channel is
\begin{align}\label{equ1_2_27}
  \textbf{h}_{CDD-OFDM}
   =  \{ &[h^0(0),...,h^0(L-1), \underbrace{0,...,0}_{\delta-L}], [h^{1}(0),..., h^{1}(L-1), \underbrace{0, ..., 0}_{\delta-L}], \nonumber \\
   &... [h^{N_t-1}(0),..., h^{N_t-1}(L-1), \underbrace{0, ...,0}_{\delta-L} ] ,0, ..., 0 \},
\end{align}
which has the same form as the equivalent ISI channel of CDD-V-OFDM (\ref{equ1_2_16}).

Despite the similarities between the CDD-OFDM and CDD-V-OFDM systems, differences are obvious. Our CDD-V-OFDM system can accommodate as many as $\lfloor \frac{NK}{L} \rfloor$ transmit antennas, while in the CDD-OFDM, the maximum number of antenna is $\lfloor \frac{N}{L} \rfloor$  in order to have that all cyclically shifted channel coefficients are not overlapped and thus achieve the full spatial and multipath diversities. Another difference between the CDD-V-OFDM and the CDD-OFDM systems is on cyclic delay shift. Here we give a proposition about the relationship between the cyclic shifts of the CDD-V-OFDM and CDD-OFDM systems.
\newtheorem{lemma}{Proposition}
\begin{lemma}
    Let $\overline{\textbf{x}} = [\overline{x}(0), \overline{x}(1), ..., \overline{x}(NK-1)]$ be a signal in time domain. After cyclic shifts in OFDM and V-OFDM systems with shift amounts $\delta_m$ and $\overline{\delta}_m$, respectively, we get the cyclically shifted signals as $\overline{\textbf{x}}_{OFDM}^m = [\overline{x}^m(0), \overline{x}^m(1), ..., \overline{x}^m(NK-1)] = [\overline{x}(\delta_m), ..., \overline{x}(NK-1), ..., \overline{x}(0), ..., \overline{x}(\delta_m-1)]$, and $\overline{\textbf{x}}_{V-OFDM}^m = [(\overline{\textbf{x}}^m(0))^T, (\overline{\textbf{x}}^m(1))^T, ..., (\overline{\textbf{x}}^m(N-1))^T] = [(\overline{\textbf{x}}(\overline{\delta}_m))^T, \\ ..., (\overline{\textbf{x}}(N-1))^T, (\overline{\textbf{x}}(0))^T, ..., (\overline{\textbf{x}} (\overline{\delta}_m-1))^T ]$, respectively, where $\overline{\textbf{x}}(t) = [\overline{x}_0(t), \overline{x}_1(t), ..., \overline{x}_{K-1}(t)]^T$, and $\overline{x}_{k}(t) = \overline{x}(k+tK)$. The two kinds of cyclic shifts have the same output signal, i.e., $\overline{\bf{x}}^m_{OFDM}= \overline{\bf{x}}^m_{V-OFDM}$ for any signal $\overline{\bf x}$, if and only if $\delta_m = K\overline{\delta}_m$.
\end{lemma}
\begin{IEEEproof}
    We first prove the sufficiency. In the cyclically shifted V-OFDM signal $\overline{\textbf{x}}_{V-OFDM}^m$, we consider the $k$th symbol of the $t$th polyphase
    \begin{align}\label{equ1_2_27}
        \overline{x}_k^m(t)
        =&\overline{x}_k(t+\overline{\delta}_m) \nonumber \\
        =&\overline{x}\left( ((k+ K(t + \overline{\delta}_m)))_{KN} \right) \nonumber \\
        =&\overline{x}\left( ((k+ Kt + K\overline{\delta}_m))_{KN} \right) \nonumber \\
        =&\overline{x}\left( ((k+ Kt + \delta_m))_{KN} \right) \nonumber \\
        =&\overline{x}^m \left( k+ Kt \right),
    \end{align}
    which is the $(k+Kt)$th symbol of the cyclically shifted OFDM signal $\overline{\textbf{x}}_{OFDM}^m$. The sufficiency is proved. We next prove the necessity. When $\overline{\textbf{x}}_{OFDM}^m = \overline{\textbf{x}}_{V-OFDM}^m$ holds. For any $\overline{\textbf{x}}$, consider the $(k+Kt)$th symbols of $\overline{\textbf{x}}_{OFDM}^m$ and $\overline{\textbf{x}}_{V-OFDM}^m$, we know that $\overline{x}^m \left( k+ Kt \right) = \overline{x}_k^m(t)$. Furthermore, $\overline{x}^m \left( k+ Kt \right) = \overline{x}\left( ((k+ Kt + \delta_m))_{KN} \right)$ and $\overline{x}_k^m(t)$ are as follows,
    \begin{align}\label{equ1_2_28}
        \overline{x}_k^m(t)
        =&\overline{x}_k(t+\overline{\delta}_m) \nonumber \\
        =&\overline{x}\left( ((k+ K(t + \overline{\delta}_m)))_{KN} \right) \nonumber \\
        =&\overline{x}\left( ((k+ Kt + K\overline{\delta}_m))_{KN} \right).
    \end{align}
    Then, the equation $\overline{x}^m \left( k+ Kt \right) = \overline{x}\left( ((k+ Kt + \delta_m))_{KN} \right) = \overline{x}_k^m(t) = \overline{x}\left( ((k+ Kt + K\overline{\delta}_m))_{KN} \right)$ holds for all the input signal $ \overline{x}(n)$. This implies that $\delta_m = K\overline{\delta}_m$.
\end{IEEEproof}

We next show an example on how the equivalent ISI channel changes in the CDD-V-OFDM system. Suppose $K=2$, $N=8$, $N_t = 2$, and $L=4$. The original CIR from the two transmit antennas are
\begin{eqnarray}
  \textbf{h}^0 &=& [h^0(0), h^0(1), h^0(2), h^0(3), 0, 0, 0, 0], \\
  \textbf{h}^1 &=& [h^1(0), h^1(1), h^1(2), h^1(3), 0, 0, 0, 0].
\end{eqnarray}
The blocked vector channels for the first antenna are
\begin{eqnarray}
  \textbf{h}^0_0 &=& [h^0(0), h^0(2), 0, 0, 0, 0, 0, 0], \\
  \textbf{h}^0_1 &=& [h^0(1), h^0(3), 0, 0, 0, 0, 0, 0].
\end{eqnarray}
and the corresponding z-domain matrix is
\begin{equation}
  \mathcal{H}_{eqv}^0(z)
  =
  \begin{bmatrix}
  H_{eqv_0}^0(z) & z^{-1} H_{eqv_1}^0(z) \\
  H_{eqv_1}^0(z) & H_{eqv_0}^0(z)
  \end{bmatrix},
\end{equation}
where
\begin{eqnarray}
  H_{eqv_0}^0(z) &=& h^0(0)+z^{-1}h^0(2), \\
  H_{eqv_1}^0(z) &=& h^0(1)+z^{-1}h^0(3).
\end{eqnarray}
For the second antenna, after the cyclic shift with amount $\overline{\delta}_m = 2$, we have
\begin{eqnarray}
  \textbf{h}^1_0 &=& [0, 0, h^1(0), h^1(2), 0, 0, 0, 0], \\
  \textbf{h}^1_1 &=& [0, 0, h^1(1), h^1(3), 0, 0, 0, 0].
\end{eqnarray}
The corresponding z-domain matrix is
\begin{equation}
  \mathcal{H}_{eqv}^1(z)
  =
  \begin{bmatrix}
  H_{eqv_0}^1(z) & z^{-1} H_{eqv_1}^1(z) \\
  H_{eqv_1}^1(z) & H_{eqv_0}^1(z)
  \end{bmatrix},
\end{equation}
where
\begin{eqnarray}
  H_{eqv_0}^1(z) &=& z^{-2}h^1(0)+z^{-3}h^1(2), \\
  H_{eqv_1}^1(z) &=& z^{-2}h^1(1)+z^{-3}h^1(3),
\end{eqnarray}
Then, the $0$th and the $1$th polyphase components of the equivalent channel are the sum of the cyclically shifted channels from the two transmit antennas, which are
\begin{eqnarray}
  \textbf{h}_{eqv0} &=& [h^0(0), h^0(2), h^1(0), h^1(2), 0, 0, 0, 0], \\ 
  \textbf{h}_{eqv1} &=& [h^0(1), h^0(3), h^1(1), h^1(3), 0, 0, 0, 0],    
\end{eqnarray}
respectively. Because we make a cyclic shift for the V-OFDM signal with amount $\overline{\delta}_m = 2$, the original channel $\textbf{h}^1$ will be equivalently shifted with amount $K\overline{\delta}_m = 4$, and we can obtain the equivalent ISI SISO channel as
\begin{align}
    \textbf{h}_{eqv} = [h^0(0), h^0(1), h^0(2), h^0(3), h^1(0), h^1(1),h^1(2), h^1(3), \underbrace{0, ... , 0}_8],
\end{align}
which will be the same as the equivalent ISI channel of the CDD-OFDM of the two transmit antennas with cyclic shift amount $4$. The equivalent z-domain pseudocirculant matrix for the V-OFDM is
\begin{equation}
  \mathcal{H}_{eqv}(z)
  =
  \begin{bmatrix}
  H_{eqv_0}(z) & z^{-1} H_{eqv_1}(z) \\
  H_{eqv_1}(z) & H_{eqv_0}(z)
  \end{bmatrix},
\end{equation}
where the $0$th and the $1$th polyphase components of the equivalent channel in z-domain are
\begin{align}
   H_{eqv_0}(z) &= h^0(0)+z^{-1}h^0(2)+z^{-2}h^1(0)+z^{-3}h^1(2), \\ 
   H_{eqv_1}(z) &= h^0(1)+z^{-1}h^0(3)+z^{-2}h^1(1)+z^{-3}h^1(3),    
\end{align}
respectively.

Let us see an example of the capability of accommodating antennas in the CDD-V-OFDM scheme in LTE system. Let the DFT size be $N = 64$, channel length be $L = 16$, CP length be $\Gamma = 16$, the number of the transmit antennas be $N_t = 4$. The multipath fading channel from the $m$th transmit antenna is
\begin{equation}
    \textbf{h}^m = [h^m(0), h^m(1), ..., h^m(15)]^T, m=0, 1, 2, 3.
\end{equation}
In CDD-OFDM system, with cyclic shift amount 16 for each from the second antenna, the equivalent ISI channel will be
\begin{equation}
    \textbf{h}_{CDD-OFDM} = [(\textbf{h}^0)^T, (\textbf{h}^1)^T, (\textbf{h}^2)^T, (\textbf{h}^3)^T],
\end{equation}
which has the size of $N = 64$. From the equivalent ISI channel, it is obvious that the multipath diversity can be fully obtained, while only $4$ transmit antennas are included in CDD-OFDM system, and the full diversity order is $64=4\times 16$. 

In our proposed CDD-V-OFDM, we take $K=2$ as an example, the CP length of the V-OFDM is 8. The blocked vector channel can be written as
\begin{align}
    \textbf{h}_{k}^m = [h^m(k), h^m(k+2), ..., h^m(k+2\cdot7)]^T, k = 0,1, m = 0, 1, ..., 7,
\end{align}
which has length $7$. Then, the OFDM signal space with DFT size $N = 64$ can accommodate as many as $8$ transmit antennas to achieve the full spatial and multipath diversities of total order $128=16\times 8$. The $k$th polyphase component of the equivalent ISI channel is
\begin{equation}
    \textbf{h}_{eqv_k} = [(\textbf{h}_{k}^0)^T, (\textbf{h}_{k}^1)^T, ..., (\textbf{h}_{k}^7)^T].
\end{equation}
The corresponding equivalent ISI channel has the size of $NK = 128$, which is
\begin{equation}
    \textbf{h}_{eqv} = [(\textbf{h}^0)^T, (\textbf{h}^1)^T, ..., (\textbf{h}^7)^T].
\end{equation}
When $K=2$, the number of the accommodating antennas in the CDD-V-OFDM is $2$ times that of the CDD-OFDM.

\subsection{Symbol-by-Symbol Detection in Each Sub-carrier After the MMSE Equalization}
\begin{figure}
  \centering
  \includegraphics[width=0.15\textwidth,angle=270]{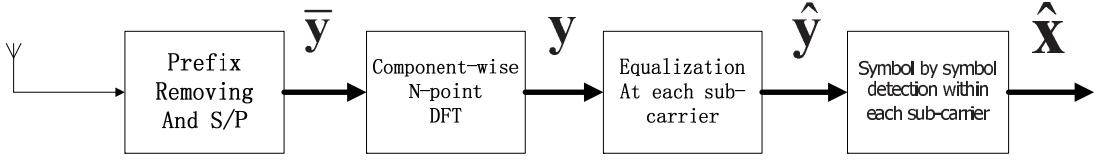}\\%
  \caption{CDD-V-OFDM receiver}\label{Fig4}
\end{figure}
As it was mentioned earlier, V-OFDM converts an ISI  channel to multiple constant MIMO subchannels and each constant MIMO subchannel includes $K$ information symbols that are ISI together by the constant MIMO channel matrix. Its ML decoding is, thus, the $K$ symbol joint ML decoding and may have a high complexity when $K$ is not small. In this subsection, we apply the MMSE equalizer/receiver for each of these constant MIMO subchannels for the CDD-V-OFDM system and the give the detection rule. The receiver of the CDD-V-OFDM system is shown in Fig. 4. The signal after removing the CP and serial-to-parallel transformation is
\begin{equation}\label{equ1_3_1}
  \overline{\textbf{y}} = [\overline{\textbf{y}}(0)^T, \overline{\textbf{y}}(1)^T, ..., \overline{\textbf{y}}(N-1)^T]^T,
\end{equation}
where $\overline{\textbf{y}}(n) = [\overline{y}_{0}(n), \overline{y}_{1}(n), ..., \overline{y}_{K-1}(n)]^T$. Take component-wise DFT transformation in the similar way as (\ref{equ1_1_8}), we have the signal in frequency domain as
\begin{equation}\label{equ1_3_2}
  \textbf{y} = [\textbf{y}(0)^T, \textbf{y}(1)^T, ..., \textbf{y}(N-1)^T]^T,
\end{equation}
where $\textbf{y}(q) = [y_{0}(q), y_{1}(q), ..., y_{K-1}(q)]^T$. Then, the relationship between the received signal and the transmitted signal in frequency domain is
\begin{equation}\label{equ1_3_3}
  \textbf{y}(q) = \textbf{H}_{eqv}(q) \textbf{x}(q) + \textbf{w}(q),
\end{equation}
where $\textbf{w}(q)$ is the equivalent noise in the frequency domain, and the equivalent channel matrix for the $q$th sub-carrier is
\begin{equation}\label{equ1_3_4}
  \textbf{H}_{eqv}(q) = \textbf{U}_q^H \overline{\textbf{H}}_{eqv}(q) \textbf{U}_q,
\end{equation}
where $\textbf{U}_q$ is a unitary matrix with entry at $s$th row and $m$th column as
\begin{equation}\label{equ1_3_5}
  [\textbf{U}_q]_{s,m} = \frac{1}{\sqrt{K}} \exp(-\frac{2 \pi (q + s N)}{NK}),
\end{equation}
where $s,m = 0, 1, ..., K-1$. The $\overline{\textbf{H}}_{eqv_q}$ in (\ref{equ1_3_4}) is a diagonal matrix as
\begin{equation}\label{equ1_3_6}
  \overline{\textbf{H}}_{eqv}(q) = \text{\textbf{diag}} [H_{q}, H_{q+L}, ..., H_{q+(K-1)L}],
\end{equation}
where
\begin{equation}\label{equ1_3_7}
  H_{l} = \sum\limits_{m=0}^{NK-1} h_{eqv}(m) \exp(-\frac{j 2 \pi m l}{NK}).
\end{equation}
With the frequency domain equivalent $\textbf{H}_{eqv}(q)$, we have the MMSE equalizer for the $q$th sub-carrier as
\begin{equation}\label{equ1_3_8}
  \textbf{W}_{q}^{MMSE} = (\textbf{H}_{eqv}(q)^H \textbf{H}_{eqv}(q) + \rho^{-1} \textbf{I})^{-1} \textbf{H}_{eqv}(q)^H.
\end{equation}
The signal after equalization is
\begin{equation}\label{equ1_3_9}
  \hat{\textbf{y}} = [\hat{\textbf{y}}(0)^T, \hat{\textbf{y}}(1)^T, ..., \hat{\textbf{y}}(N-1)^T]^T,
\end{equation}
where
\begin{equation}\label{equ1_3_10}
  \hat{\textbf{y}}(q) =\textbf{W}^{MMSE}(q) \textbf{y}(q), q = 0, 1, ..., N-1,
\end{equation}
and $\hat{\textbf{y}}(q) = [\hat{y}_{0}(q), \hat{y}_{1}(q), ..., \hat{y}_{N-1}(q)]^T$. Then, the detection is processed symbol by symbol, which is
\begin{equation}\label{equ1_3_11}
  \hat{x}_k(q) = \arg \min_{i} \| C_{qK+k} x(i) - \hat{y}_k(q) \|^2,
\end{equation}
where $x(i)$ is all the possible signal for $\hat{x}_q(k)$ and
\begin{equation}\label{equ1_3_12}
  C_{qK+k} = \frac{1}{K} \sum\limits_{k=0}^{K-1} \frac{\mid H_{q+kN} \mid^2}{\mid H_{q+kN} \mid^2 + \rho^{-1}}.
\end{equation}
Then the estimated signal is
\begin{equation}\label{equ1_3_13}
  \widehat{\textbf{x}} = [\widehat{\textbf{x}}(0)^T, \widehat{\textbf{x}}(1)^T, ..., \widehat{\textbf{x}}(N-1)^T]^T,
\end{equation}
where $\widehat{\textbf{x}}(q) = [\hat{x}_{0}(q), \hat{x}_{1}(q), ..., \hat{x}_{K-1}(q)]$. The detection computational complexity is only linear proportional to the vector length $K$ and the DFT size $N$.

\subsection{Performance Analysis}
In this sucsection, based the detection rule (\ref{equ1_3_12}) and the MMSE equalization in frequency domain, we analyze the diversity order of the CDD-V-OFDM system. Based on the derivation in Seciton \Rmnum{3}.B, we directly apply the results obtained in \cite{li2012performance} to get the diversity order.

Assume that the CIRs from different antennas are known at the receiver and the condition (\ref{equ1_2_24}) holds for the cyclic shift amounts. Let $R$ be the transmission data rate, $L$ be the length of the CIR from a single transmit antenna, $N_t$ be the number of the transmit antennas and $K$ be the vector length of the signals in the CDD-V-OFDM system. Then, following \cite{li2012performance}, the diversity order of the CDD-V-OFDM system is
\begin{equation}\label{eqv4_1}
    d_{\text{CDD-V-OFDM}}^{\text{MMSE}} = \min \{ \lfloor 2^{-R}K \rfloor, N_t L \} +1.
\end{equation}
\textbf{Remarks:}
\begin{itemize}
  \item When $\lfloor 2^{-R}K \rfloor< N_t L$ holds, the diversity order of the CDD-V-OFDM is determined by $2^{-R}K$. Under this condition, if $R$ decreases or $K$ increases, we can obtain a higher diversity order.
  \item When $\lfloor 2^{-R}K \rfloor> N_t L$ holds, the diversity order of the CDD-V-OFDM is determined by $N_t L$. Increasing the number of transmit antennas $N_t$ can improve the diversity order of the CDD-V-OFDM system.
  \item The V-OFDM system is the special case of the CDD-V-OFDM system when $N_t = 1$. Thus, the diversity order of the V-OFDM is
    \begin{equation}\label{eqv4_2}
        d_{\text{V-OFDM}}^{\text{MMSE}} = \min \{ \lfloor 2^{-R}K \rfloor, L \} +1,
    \end{equation}
    From (\ref{eqv4_1}) and (\ref{eqv4_2}), we know that the the CDD-V-OFDM system has the same diversity order as the V-OFDM when $\lfloor 2^{-R}K \rfloor< L$. But the diversity order of the V-OFDM system can not be improved when $\lfloor 2^{-R}K \rfloor > L$, while the proposed CDD-V-OFDM can improve the diversity order by increasing the number of the transmit antennas.
  \item The CDD-OFDM is also the special case of the CDD-V-OFDM when $K=1$. The diversity gain can not be fully obtained unless the error correction coding is used over all the subcarriers, which can be seen from the simulation results. To achieve the maximal spatial and multipath diversities, the maximum number of accommodating antennas in the CDD-OFDM system is $\lfloor N/L \rfloor$, while in the CDD-V-OFDM system, higher diversity order is obtained with the maximum number of the transmit antennas as $\lfloor NK/L \rfloor$. Furthermore, the CP length of each polyphase in the CDD-V-OFDM, i.e., $\lceil L/K \rceil$, is only $1/K$ that of the CDD-OFDM, i.e., $L$.
\end{itemize}
\begin{figure}
  \includegraphics[width=0.9\textwidth]{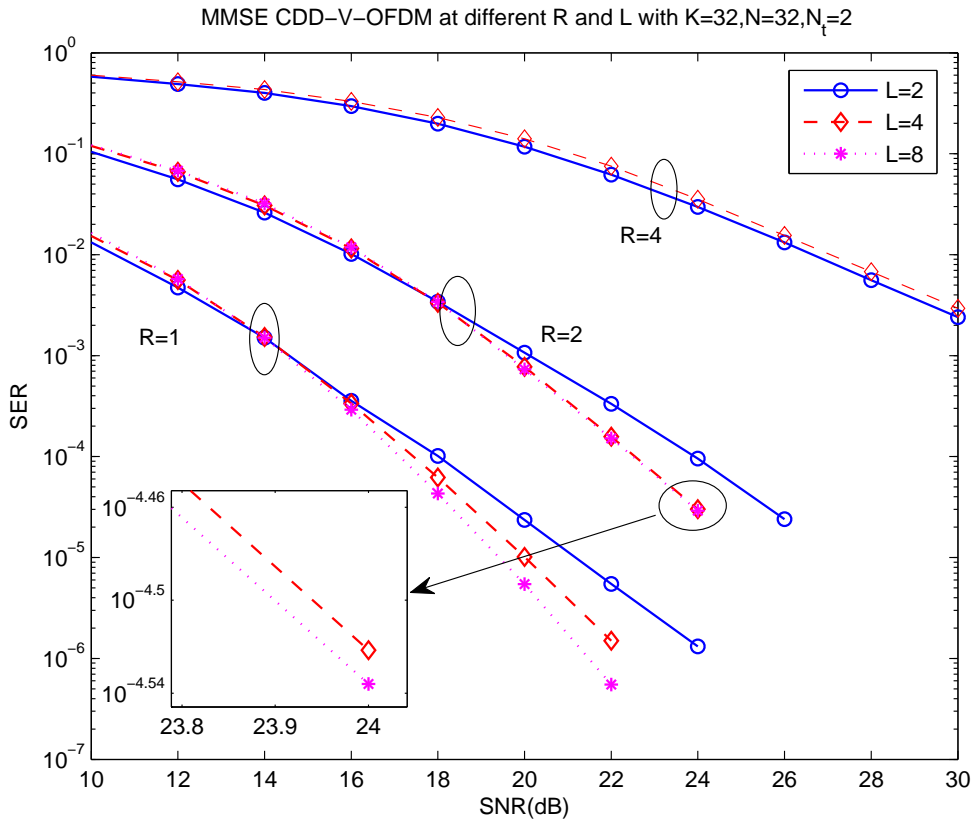}\\
  \caption{MMSE CDD-V-OFDM at different $R$, and $L$ with $K=32$, $N=32$, $N_t=2$}\label{simu_fig1}
\end{figure}
\begin{figure}
  \includegraphics[width=0.9\textwidth]{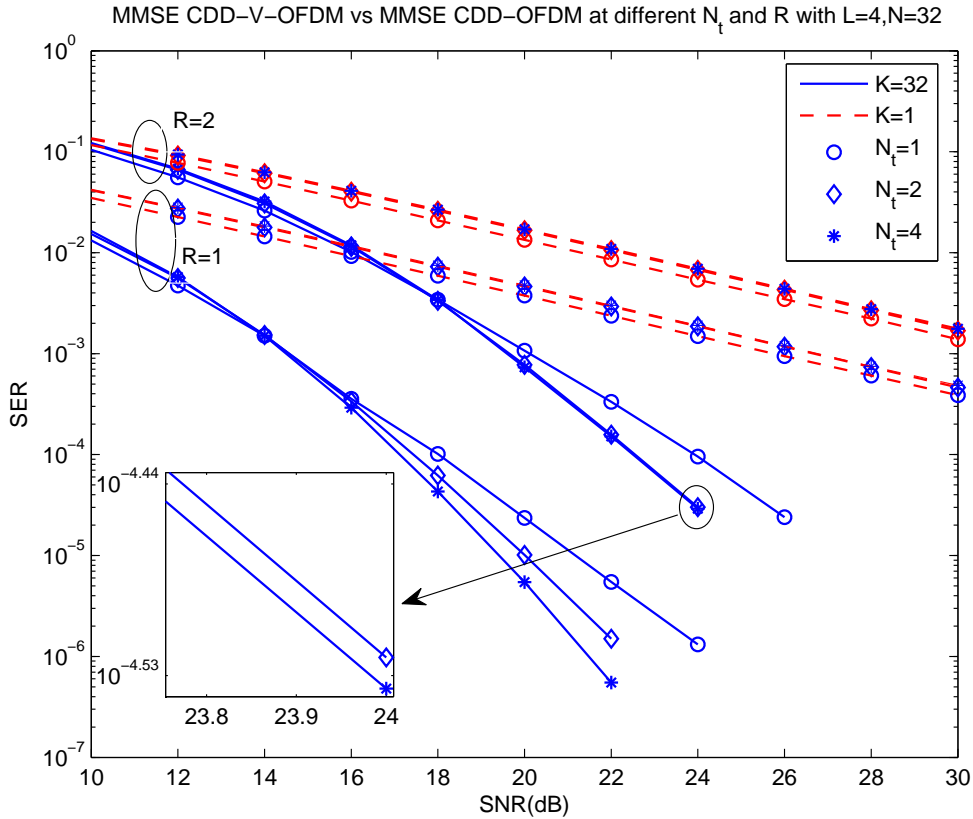}\\
  \caption{MMSE CDD-V-OFDM vs MMSE CDD-OFDM at different $N_t$, and $R$ with $L=4$, $N=32$}\label{simu_fig2}
\end{figure}
\begin{figure}
  \includegraphics[width=0.9\textwidth]{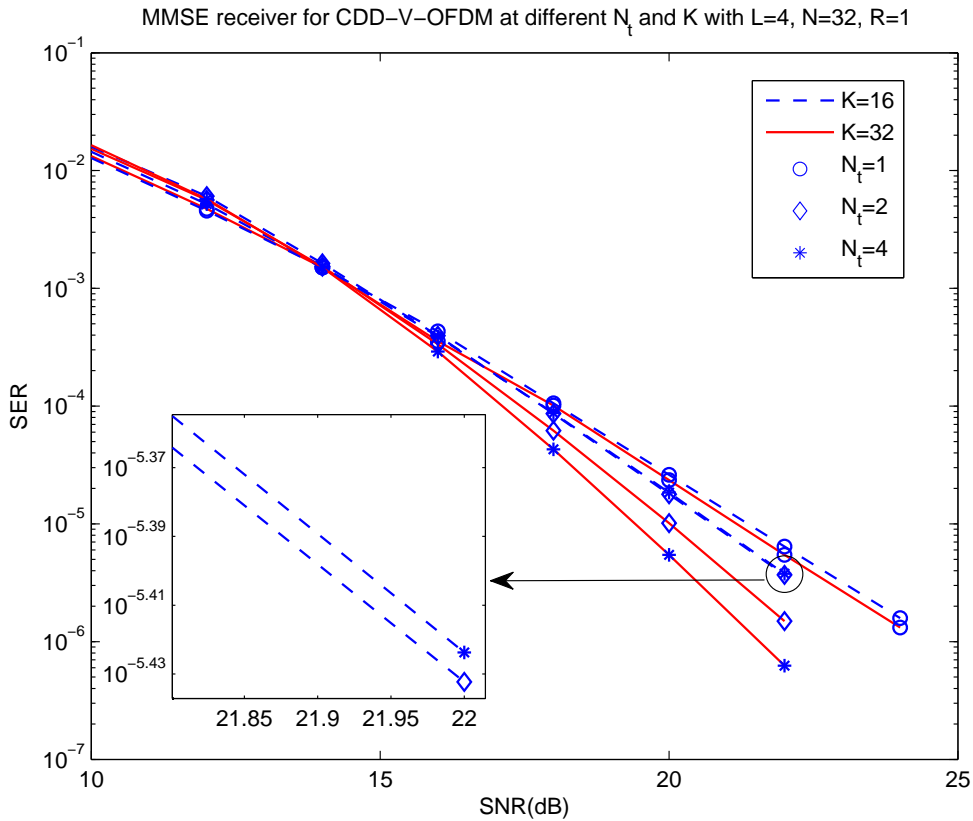}\\
  \caption{MMSE receiver for CDD-V-OFDM at different $N_t$, and $K$, with $L=4$, $N=32$, $R=1$}\label{simu_fig3}
\end{figure}
\begin{figure}
  \includegraphics[width=0.9\textwidth]{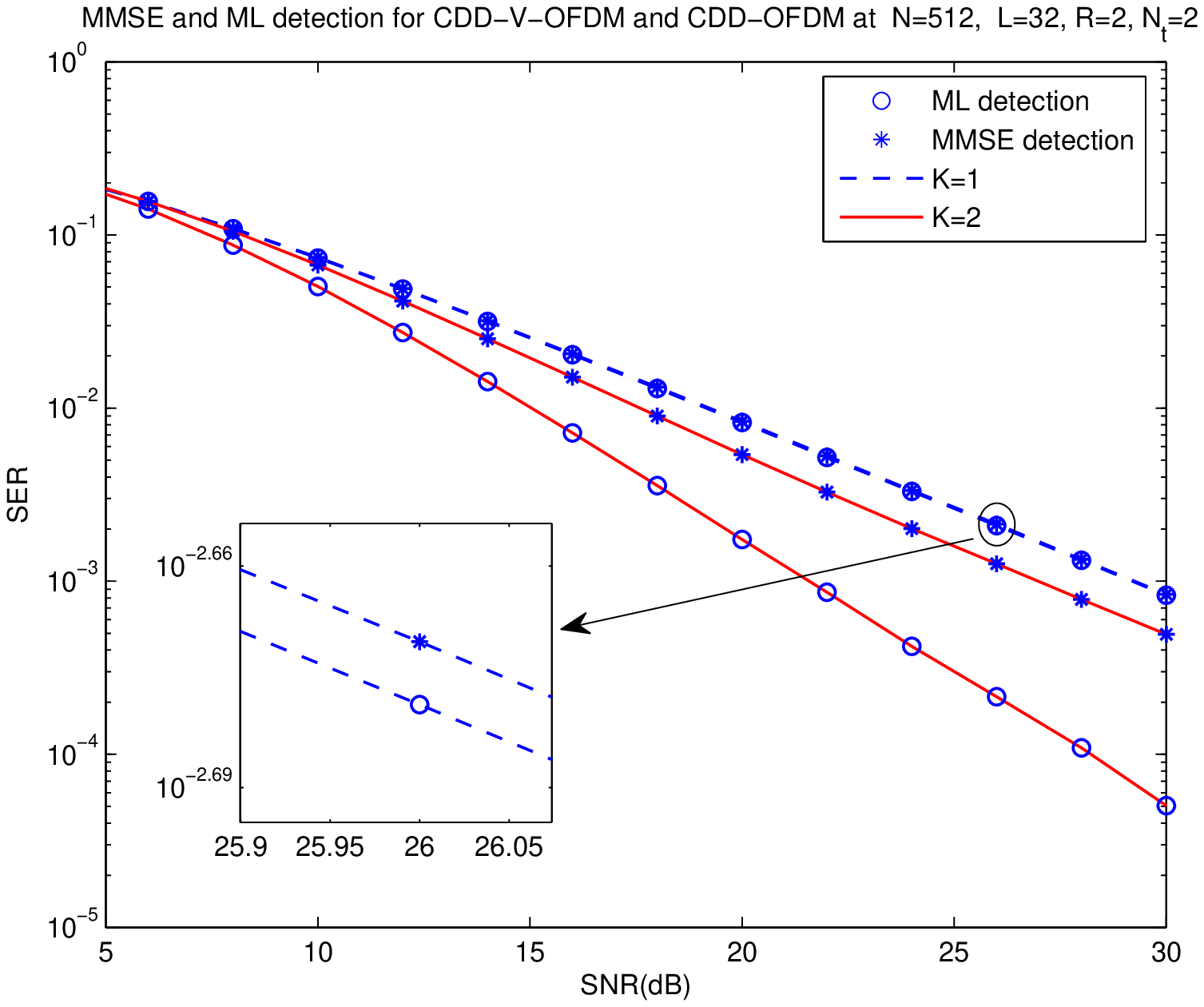}\\
  \caption{MMSE and ML detection for CDD-V-OFDM and CDD-OFDM at $N=512$ , $L=32$, $R=2$, $N_t=2$}\label{simu_fig4}
\end{figure}
\section{Simulation Results}
In this section, we present some simulation results for the symbol error rate (SER) performance to show the diversity orders of the CDD-V-OFDM systems. Since the CDD-V-OFDM system contains multiple transmit antennas, we let the sum power of all the transmit antennas of the CDD-V-OFDM be $P_{\text{CDD-V-OFDM}}$. Then, we analyze the diversity orders of the systems under the condition $P_{\text{CDD-V-OFDM}} = 1$. In all the simulations, the fading channel is i.i.d. complex Gaussian distributed with zero mean and unit variance. For all the figures, the systems are uncoded. The cyclic prefix length is chosen as $\Gamma = \lceil L/K \rceil+1$ for the CDD-V-OFDM system.

Fig. 4 shows the SER performance of the MMSE CDD-V-OFDM system at different $R$ and $L$ with $N_t=2$, $N=32$ and $K=32$. For the case $R = 4$, $2^{-R}K = 2$, and $L$ increasing from $2$ to $4$, $2^{-R}K < N_t L$ always holds. Then, the diversity order is always $d_{\text{CDD-V-OFDM}}^{\text{MMSE}} = 5$, which can be seen from the slopes of the curves. In the case $(R,K)=(2,32)$, $2^{-R}K = 8$, and the condition $2^{-R}K > N_t L$ still holds, then the diversity order of the CDD-V-OFDM is $d_{\text{CDD-V-OFDM}}^{\text{MMSE}} = N_t L +1= 2 \times 2+1 = 5$. When $L=4$ and $8$, the condition $2^{-R}K \leq N_t L$ holds again, and the corresponding diversity order is $d_{\text{CDD-V-OFDM}}^{\text{MMSE}} = N_t L + 1= 9$. In the case $R=1$, $2^{-R}K = 16$. When $L=2, 4, 8$, $N_t L = 4, 8, 16$, respectively, the condition $2^{-R}K \geq N_t L$ holds, and the diversity order of the CDD-V-OFDM is $d_{\text{CDD-V-OFDM}}^{\text{MMSE}} = 5, 9, 17$, respectively. This shows that under the condition $\lfloor 2^{-R}K \rfloor \geq N_t L$, the diversity order of the CDD-V-OFDM system increases along the increase of $L$. Fig. 4 also shows that under the condition $2^{-R}K \leq N_t L$, the diversity order of the CDD-V-OFDM system increases along the decrease of $R$. From $(R,L) = (4,8)$ to $(R,L) = (2,8)$ and $(R,L) = (1,8)$, $N_t L = 16$, and the condition $2^{-R}K \leq N_t L$ always holds. Then the diversity order of the system increases from $d_{\text{CDD-V-OFDM}}^{\text{MMSE}} = 5$ to $9$ and $17$, respectively. This result is shown in Fig. 4, which matches the result (\ref{eqv4_1}).

Fig. 5 shows the SER performance of the MMSE CDD-V-OFDM at different $N_t$ and $R$ with $L=4$ and $K=32$. In Fig. 5, the special case $K=1$ is also shown, which is the CDD-OFDM system. For the CDD-OFDM system, the diversity order is always $d_{\text{CDD-OFDM}}^{\text{MMSE}} = 1$, whenever $N_t$ increases from $1$ to $2$ and $4$. The maximum number of transmit antennas in the CDD-OFDM system is $N_t = 8$. The CDD-V-OFDM system always obtains higher diversity order than the CDD-OFDM. When $(R,N_t) = (2,1)$, $d_{\text{CDD-V-OFDM}}^{\text{MMSE}} = N_t L +1=5$. When $(R,N_t) = (2,2)$, and $(2,4)$, $d_{\text{CDD-V-OFDM}}^{\text{MMSE}} = 9$. In the case $R=1$, when $N_t$ increases from $1$ to $2$ and $4$, $d_{\text{CDD-V-OFDM}}^{\text{MMSE}} = 5$, $9$, and $17$, respectively. 

In Fig. 6, the SER performance of the MMSE CDD-V-OFDM system is presented at different $R$ and $N_t$ with $K=16$ and $32$, $L=4$. In the case of $K = 16$, when the number of the transmit antennas increases from $N_t=1$ to $2$, the diversity order is increased from $d_{\text{CDD-V-OFDM}}^{\text{MMSE}} =5$ to $9$. When $N_t = 4$, the condition $\lfloor 2^{-R}K\rfloor > N_t L$ holds, and then the diversity order is also $d_{\text{CDD-V-OFDM}}^{\text{MMSE}} =9$. In the case $K=32$, the condition $\lfloor 2^{-R}K\rfloor <N_t L$ always holds, then $d_{\text{CDD-V-OFDM}}^{\text{MMSE}}$, the diversity of the system increases from $5$ to $9$ and $17$ along $N_t=1, 2, 4$, respectively. Fig. 6 verifies that the increase of $K$ can improve the diversity order of the system.

In Fig.7, $N=512$, $L=32$, $N_t = 2$ and $R=2$. Due to the complexity of ML detection, it is not easy to simulate for the cases with large $K$ and thus small $K=1$ and $2$ are simulated. In this case, the condition $\lfloor 2^{-R}K \rfloor <N_t L $ always holds. Then, $d_{\text{CDD-V-OFDM}}^{\text{MMSE}} = \lfloor 2^{-2}2 \rfloor +1 = 1$.
In comparison with the special case $K=1$, i.e., the CDD-OFDM system, it is obvious that the CDD-V-OFDM has better performance.


\section{Conclusion}
In this paper, we have proposed a CDD transmission for the single antenna V-OFDM system when there are multiple transmit antennas. With our proposed CDD-V-OFDM system, $K$ times more transmit antennas can be accommodated than  the conventional CDD-OFDM system to collect both spatial and multipath diversities, which will become more interesting for a massive MIMO system, where $K$ is the vector size in V-OFDM. With the MMSE linear receiver applied to each subchannel, by following the previous result in \cite{li2012performance} for a V-OFDM system, we have obtained the diversity order for our proposed  CDD-V-OFDM system:  $d_{\text{CDD-V-OFDM}}^{\text{MMSE}} = \min \{ \lfloor 2^{-R}K \rfloor, N_t L \} +1$, where $R$ is the transmission rate,  $N_t$ is the number of transmit antennas, and $L$ is the ISI channel length from a transmit antenna to a receive antenna.


%

\appendices
%
%
%
%
%

\ifCLASSOPTIONcaptionsoff
  \newpage
\fi

\end{document}